\documentclass[fleqn,twoside]{article}
\usepackage[headings]{espcrc2}
\usepackage[dvips]{epsfig}
\usepackage{graphicx}
\usepackage{amssymb}
\usepackage[figuresright]{rotating}

\newcommand{\Bb}{J/\psi}
\newcommand{\ptb}{P_T^b}
\newcommand{\mm}  {\mu^+\mu^-}
\def \to {\rightarrow}

\title{
Prospects for b-quark production cross 
section measurements in pp collisions at the LHC
}
\author{A. Sherstnev
\address{Cavendish Laboratory, University of Cambridge, Cambridge, UK}%
}

\begin{document}

\begin{abstract}
A brief review of theoretical and experimental aspects of $b$-quark 
production measurements at the LHC. 
\vspace{1pc}
\end{abstract}
\maketitle

\section{Introduction}
Study of heavy quarks will be an important research area in experiments at 
the LHC. Investigation of heavy quarks is interesting on its own (CP-violation, 
properties of $B$-hadrons, constrains on hadronization models, heavy flavour 
PDFs, etc.) and as a background to other processes (Higgs, top quark, Beyond 
SM particle production). 

One of the first problems in the research will be measurement of the total 
cross section $\sigma_{bb}$. Due to the huge cross section, large statistics 
will be available at once after the collider launch. So the measurements 
will help to check and understand the LHC detectors. 

\section{Theoretical methods}
Heavy quarks introduce additional complications in calculations 
due to non-zero masses, which play the role of supplementary energy scales. 
The 15 year story of the $b$-quark production cross section calculations and 
measurements at the Tevatron shows a remarkable example of the troubles 
(see details in a nice review~\cite{Mangano:2004xr}). 

All approaches to the calculations can be divided into two classes: the first 
class consists of cross section calculators, producing total cross sections 
(sometimes within cuts applied) and kinematic distributions: NLO, ACOT, BSMN, 
FONLL, GM-VFNS (see~\cite{Alekhin:2005dy} and refs. within). These methods are 
able to take into account radiative corrections and/or resummation effects (e.g. 
large logarithms of $P_T/m_b$ or $\hat{S}/m_b$). 

The state of the art amongst the approaches in the $\sigma_{b\bar{b}}$ calculations 
is FONLL~\cite{Cacciari:1998it}. The method is based on matching 
of the NLO calculations and resummation of $\log{\ptb/m_b}$. Due to the high 
$\ptb$ region available at the Tevatron, we have 2 different energy scales, 
$m_b$ and $\ptb$, in the process $pp\to b\bar{b}$. So large logarithms 
$\log{\ptb/m_b}$ appear in all orders of the QCD expansion of $M(pp\to b\bar{b})$. 
FONLL resums the terms in the NLL approximation and uses the following prescription: 
$$
\rm\sigma_{fonll} = \sigma_{nlo}+(\sigma_{rs}-\sigma_{nlo,m_b\to0}))G({\it P_T})\; .
$$
In order to exclude double counting a massless approximation to NLO should be 
subtracted from the resummation contribution. The weight function G($P_T$) 
ensures a proper application region for the resummation (here, $\ptb>5m_b$). 
The resummation term has the structure: 
$$
\sigma_{rs} = f_i(x_1,\mu)f_j(x_2,\mu)\sigma_{ij\to k}(\hat{s},\mu) D_{k\to b}(\mu,\mu_0)\; ,
$$
where $\mu \sim P_T$ and $\mu_0 \sim m_b$ are factorization scales. The functions 
here are convoluted with respect to momentum fractions. $\sigma_{ij\to k}$ 
is an NLO expression with no large logarithms thanks to the choice $\mu$ 
(it is calculated in pQCD). The function $D(\mu,\mu_0)$ describes the resummed 
final-state collinear logarithms. Due to the large $m_b\gg\Lambda_{QCD}$, 
initial expressions for $D(\mu_0,m)$ can be calculated perturbatively. 
The method produces total cross sections and distributions. It has been 
realized for heavy quark hadroproduction~\cite{Cacciari:1998it} and 
photoproduction~\cite{Cacciari:2001td}. 

\begin{figure}[t]
\begin{center}
\epsfxsize=7.5cm
\epsfysize=4.4cm
\epsffile{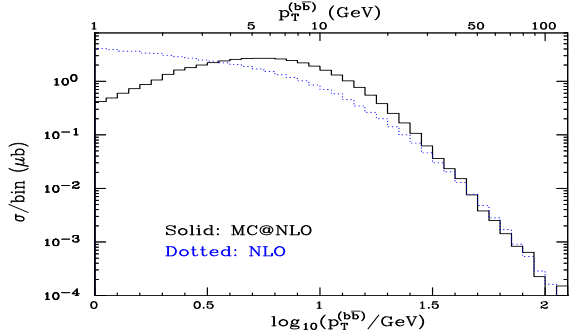}
\end{center}
\caption{
$P_T$ of $b$-quarks in MC@NLO and pure NLO approaches~\cite{Frixione:2003ei}.
}
\label{fg:mcatnlo-pt}
\end{figure}

\begin{figure}[t]
\begin{center}
\epsfxsize=7.5cm
\epsfysize=4.4cm
\epsffile{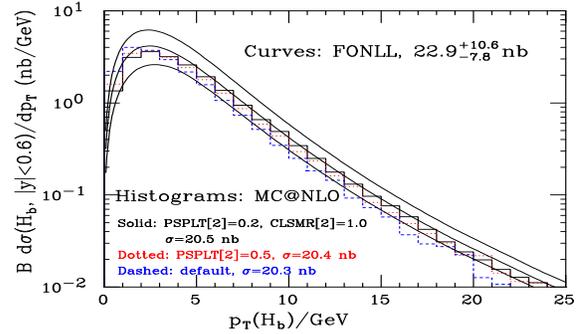}
\end{center}
\caption{
Comparison of $P_T^B$ in MC@NLO and FONLL at the Tevatron~\cite{Cacciari:2003uh}. 
}
\label{fg:mcatnlo-fonll}
\end{figure}

The second class of codes is Monte-Carlo generators, which can produce 
Monte-Carlo events. The state-of-the-art player for the task considered is 
MC@NLO~\cite{Frixione:2002ik}. It uses an exact NLO calculator to generate 
events at NLO and a special procedure to match the events with further 
parton showers. So the program can give fully exclusive events at NLO with 
showering, hadronization and decay effects. The HERWIG showering and 
hadronization model is applied. The big advantage of the code is fast event 
generation ($\sim$10 times faster a LO codes). The program gives 
reliable predictions in the whole $\ptb$ region due to the large QCD scales 
used, $Q^2_{ren}=Q^2_{fact}=m_b^2+P_T^2$. In addition, variation of the scales 
allows one to estimate NNLO uncertainties. Absence of massive $b$-quarks in the 
initial state simplifies the use of PDFs. Fig.~\ref{fg:mcatnlo-pt} shows the 
impact of the showering effects on the $\ptb$ distribution. 
Comparison (see Fig.~\ref{fg:mcatnlo-fonll}) shows very good agreement 
between the FONLL and MC@NLO approaches.

Theorists and Monte-Carlo code authors have 2-3 years until real data at 
the LHC. So, a couple of problems could be solved. {\it 1)} NNLO calculations 
for $\sigma_{bb}$. QCD NLO results have a big scale dependence: variation 
$0.5<Q_{ren}/Q_{fac}<2$ gives $\sigma_{bb}$ = 276 -- 645 $\mu$b with a central 
value of 496 $\mu$b. Due to recent progress in QCD NNLO calculations this 
task can be posed and attempted. {\it 2)} Matching NLO with resummation 
contributions in events: Tevatron experience shows the importance of large 
logarithms in the high $\ptb$ region. Since experimentalists require events, 
the task also has a high priority. 

\section{Tevatron: results and lessons}
There was a problem with interpretation of the $B$-hadron production cross 
section in Run I. The first value of the Data/Theory ratio was 
$\rm2.9\pm0.2\,(stat.)\pm0.4\,(syst.)$~\cite{Acosta:2001rz}, subsequent 
progress gave a new value $\rm1.7\pm0.5\,(stat.)\pm0.5\,(syst.)$~\cite{Cacciari:2002pa}. 
FONLL, new PDFs with a higher value of $\alpha_s$, and new $F_{frag}(b\to H_b)$ 
were taken into account. Run II brought new experimental improvements: better 
secondary vertex reconstruction, more efficient $B$-hadron ID, much more 
statistics available, $\ptb$ spectra down to 0 GeV. Two types of 
measurements are possible at Run II. $B$-hadron spectra in different decay 
channels in a limited $P_T$ region (0 -- 25 GeV), but with huge statistics; 
inclusive $b$-jets with the very high upper $\ptb$ limit ($\sim$ 500 GeV), 
but with lower statistics. 

\begin{figure}[t]
\begin{center}
\epsfxsize=7.5cm
\epsfysize=4.4cm
\epsffile{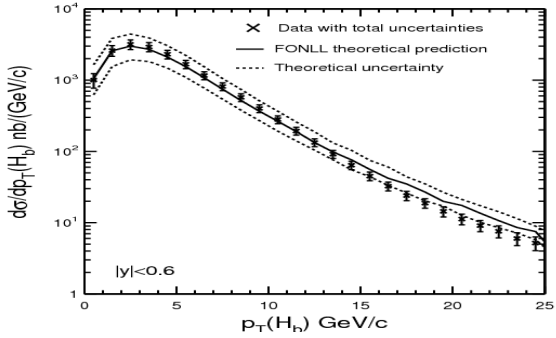}
\end{center}
\caption{
The $b$-quark $P_T$ distribution compared with FONLL. 
}
\label{fg:tevII-lowPt}
\end{figure}

$B$-hadron production measurements were published by CDF with ~40 
pb$^{-1}$~\cite{Acosta:2004yw}. Trigger: 2$\mu$ (L1) + $\Bb$ selection. 
HLT: opposite-sign muons and 2.7$<M(2\mu)<$ 4.0 GeV. Offline selection: the 
central region ($|\eta(\Bb)|<$ 0.6) and $P_T(\mu)>$ 1.5 GeV, $H_b\to\Bb$+X 
selection: pseudo proper decay length. Fig.~\ref{fg:tevII-lowPt} shows a 
comparison with theory (both FONLL and MC@NLO approaches) with cuts 
applied: $P_T(\Bb)>$ 1.25 GeV and $|\eta(\Bb)|<$ 0.6 The total cross section 
is $\rm 19.4\pm 0.3\,(stat.)\pm 0.2\,(syst.)$. Theory gives 
$\rm18.3^{+8.3}_{-5.9}\;\mu b$ (FONLL) and 17.2 $\mu b$ (MC@NLO). 

\begin{figure}[t]
\begin{center}
\epsfxsize=7.5cm
\epsfysize=4.4cm
\epsffile{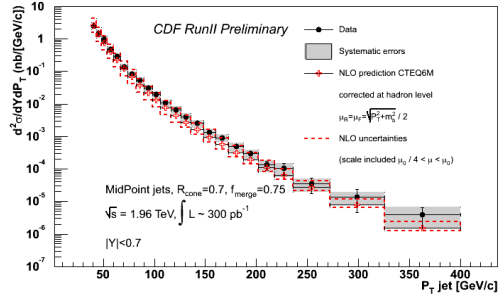}
\end{center}
\caption{
The $P_T$ spectrum of inclusive $b$-jets at the Tevatron. 
}
\label{fg:tevII-highPt}
\end{figure}

Preliminary results on high $P_T$ b-tagged jet production measurements in 
CDF are presented in~\cite{DOnofrio:2005fr}. Trigger: inclusive jet events 
with several $P_T$ thresholds (20, 50, 70, and 100 GeV). Jets are taken in 
the central region ($|\eta(j)|<$ 0.7) only. The $P_T(j)$ range is 38 -- 400 GeV. 
b-tagging is done by secondary vertex reconstruction. A comparison with 
theory is reported in Fig.~\ref{fg:tevII-highPt}. 

Two types of complementary measurements are possible on hadron machines. 
$B$-hadron spectra in the low $\ptb$ region have the advantages of high 
statistics (which means fast measurements and low statistical errors) 
and several independent channels, but the $P_T$ range is limited 
(0 -- 25 GeV). The study of inclusive $b$-jets in the high 
$\ptb$ region doesn't have this drawback, but statistical errors in the 
measurements are much higher. 

\section{CMS/ATLAS strategies}
\begin{figure}[t]
\begin{center}
\epsfxsize=7.5cm
\epsfysize=4.4cm
\epsffile{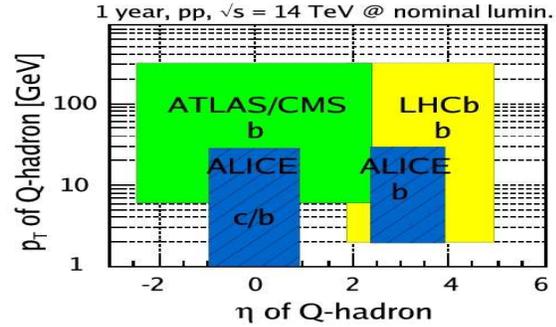}
\end{center}
\caption{
$P_T \otimes \eta$ regions available in all LHC experiments in 
heavy flavour analysis~\cite{Alekhin:2005dy}. 
}
\label{fg:lhc-pt-eta}
\end{figure}

The heavy quark production cross section is huge at the LHC, $\sigma_{b\bar{b}} \sim 500$ 
$\mu b$. So problems of triggering will be very important in all experiments. 
The most appropriate triggers in the $B$-hadron analyses are based on muons (di-muon, 
single-muon). First measurements at the LHC will be inclusive ones (at first, 
measurements of the total cross sections). Later, cross sections of selected 
channels will be available. This will provide the possibility to cross-check the 
measurements. The different acceptance of LHCb and ATLAS/CMS will allow 
complementary measurements in the detectors (see Fig.~\ref{fg:lhc-pt-eta}). 

CMS and ATLAS are general purpose detectors, optimized for high $P_T$ research tasks. 
Both detectors have precise tracking and vertex detectors, which are important 
in b-physics (more details about CMS/ATLAS may be found in~\cite{Speer:2006}). 
Due to strict conditions of triggering (ATLAS and CMS triggers should reduce the 
total rate, 40 MHz, to 200 Hz(ATLAS)/150 Hz(CMS)), there will be a competition on 
bandwidth and HLT resources for b-objects. The main triggers which will be
used in the total cross section measurements are di-muon and single-muon ones: 
ATLAS di-muon $P_T(\mu)>$ 3-4 GeV in $|\eta(\mu)|<$ 2.7 (60\% efficiency, 10Hz 
rate for $pp\to\Bb+X$ after HLT); CMS di-muon $P_T(\mu)>$ 3 GeV in $|\eta(\mu)|<$ 2.4; 
ATLAS 1-muon $P_T(\mu)>$ 8 GeV in $|\eta(\mu)|<$ 2.7 (this can help to 
reconstruct exclusive $B$-hadronic decays, but will only be applied in regimes 
with L $< 2\cdot 10^{32}$ cm$^{-2}$s$^{-1}$); CMS 1-muon 
$P_T(\mu)>$ 14 GeV in $|\eta(\mu)|<$ 2.1. 

\begin{figure}[t]
\begin{center}
\epsfxsize=7.5cm
\epsfysize=4.5cm
\epsffile{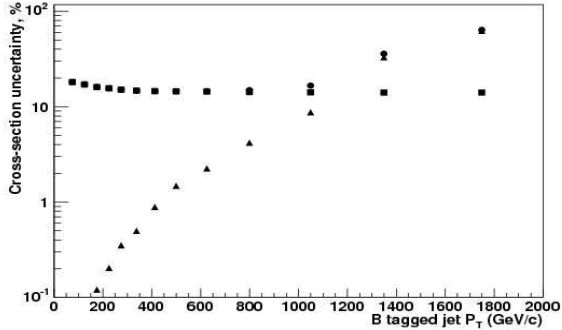}
\end{center}
\caption{
Systematic ($\blacktriangle$) and statistical ($\blacksquare$) errors in inclusive 
$b$-jet analysis in CMS~\cite{Andreev:2006}. 
}
\label{fg:cms-highPt}
\end{figure}

CMS prepared a possible strategy to measure the inclusive $b$-jet production 
cross section in the high-PT region~\cite{Andreev:2006}. $b$-jets with 
$\ptb>$ 50 GeV are analyzed. Event selection is based on the 1-muon 
L1 trigger and a HLT channel with the signature: 1 $\mu$ and 1 $b$-jet 
(secondary vertex techniques of b-tagging being used). Inclusive measurements 
are based on a fit to $P_T(\mu-j)$ ($P_T(\mu)$ with respect to the $b$-jet 
axis). The analysis estimates 3 components in the whole event sample: 
$b\bar{b}/c\bar{c}/q\bar{q}$ = 66-55\%/32-31\%/2-14\% (in the $\ptb$ range 
50 -- 1400 GeV). The main systematics in the analysis are jet energy scale 
uncertainties. 
Uncertainties depending on $\ptb$ are reported in Fig.~\ref{fg:cms-highPt}. 

\section{LHCb strategy}
LHCb is a single-arm spectrometer (the pseudorapidy range is 1.9 -- 4.9), 
optimized for $b$-quark physics (more details about LHCb are given in~\cite{Metlica:2006}). 
Due to a reduced luminosity regime (2 -- $5\cdot 10^{32}$ cm$^{-2}$s$^{-1}$ 
it has much lower $\ptb$ thresholds for $B$-hadron reconstruction (down to 
$\ptb\sim$ 1-2 GeV). Since the ``visible'' (in the detector acceptance) total 
cross section is $\sigma_{b/\bar{b}} \sim$160 $\mu$b, LHCb will produce a 
total rate of 3-$8\cdot 10^{11}$ $b\bar{b}$ pairs per year (which corresponds 
to an event rate of 30-80 kHz). LHCb trigger will provide a reduction in 
the total rate (10 MHz) down to 2 kHz. The trigger definition is crucial for 
the inclusive $b\bar{b}$ measurements. HLT trigger: 600 Hz for events with 
$\Bb\to\mm$ (trigger efficiency $\sim$75\%). 

There is no completed strategy for the total $B$-hadron production cross 
section measurement. Ideas will be taken from the Tevatron experience. The 
main strategy will be based on the $b\to\Bb\to\mm$ channel with secondary 
vertex reconstruction. Events with 3-4 muons ($b\to\mu+c$, $c\to\mu+d$) 
could be interesting for the task~\cite{Harrison:2006}. These events have 
lower backgrounds, but also lower statistics. 

\section{Conclusions}
The problem of the total $\sigma_{bb}$ measurement at the Tevatron Run I has 
resolved, FNOLL and MC@NLO equipped with new PDFs, $\alpha_{s}$ and 
re-calculated fragmentation functions describe Run II data properly. In order 
to compare the theoretical predictions with new data (Tevatron and LHC) we 
need to reduce theoretical systematics: NNLO and resummation effects in event generation. 
This sets a great challenge for theorists. Tevatron ideas on measurements of 
$\sigma_{b\bar{b}}$ suit the LHC experiments. ATLAS, CMS, LHCb have strategies, 
published or in preparation, for the task. The main advantage of the different 
geometry of ATLAS/CMS and LHCb is that the experiments will make complementary 
measurements with respect to $P_T\otimes\eta$ range. But, partly, the measurements 
could be done in the same region. This will be very important for cross-checks. 

{\bf Acknowledgments}: I thank B.~R.~Webber for useful discussions.


\end{document}